\documentclass[doublecol]{epl2} %for 2 columns style without line numbers
\usepackage{gensymb}

\title{Resistivity measurements unveil microscopic properties of CrAs}

\author{A. Nigro\inst{1,2}, P. Marra\inst{3}, C. Autieri\inst{4}, W. Wu\inst{5,6}, J. G. Cheng\inst{5,6,7}, J. Luo\inst{5,6,7} \and C. Noce\inst{1,2}}

\institute{                    
  \inst{1} Dipartimento di Fisica "E.R. Caianiello",
	Universit\'a degli Studi di Salerno, I-84084 Fisciano (SA), Italy\\
  \inst{2} CNR-SPIN, UOS di Salerno, I-84084 Fisciano (Salerno), Italy\\
  \inst{3} RIKEN Center for Emergent Matter Science, Wakoshi, Saitama 351-0198, Japan\\
  \inst{4} International Research Centre Magtop, Institute of Physics, Polish Academy of Sciences, Aleja Lotników 32/46, PL-02668 Warsaw, Poland\\
  \inst{5}Beijing National Laboratory for Condensed Matter Physics and Institute of Physics, Chinese Academy of Sciences, Beijing 100190, China\\
  \inst{6}Songshan Lake Materials Laboratory, Dongguan, Guangdong 523808, China\\
  \inst{7}School of Physical Sciences, University of Chinese Academy of Sciences, Beijing 100190, China
}

\abstract{We report resistivity measurements of a  CrAs single crystal in a wide temperature range, with the specific aim to assess the applicability of the Bloch-Gr\"uneisen formula for electron-phonon resistivity. 
We find that the resistance reaches a residual value at $T_c\sim$ 4.2 K and its temperature dependence  cannot be fitted only with a suitable Bloch-Gr\"uneisen formula in the whole temperature range, even though we are able to calculate a well defined transport Debye temperature. The observed temperature dependent resistivity seems to suggest a non phonon-mediated superconducting pairing, supporting a magnetic fluctuation mechanism as the likely glue for the superconducting coupling.}

\begin{document}

\maketitle

\section{Introduction} The recent discovery of superconductivity in chromium arsenide CrAs has attracted a lot interest~\cite{norman15} because this material has been synthesized looking for superconductivity on the verge of the antiferromagnetic order by means of the application of external pressure~\cite{wu14}. 
The superconducting transition temperature $T_c$ shows a dome-shaped pressure dependence, with a maximum $T_c\sim$ 2.17 K at a critical pressure 
$P_c$=10~kbar~\cite{wu14,kotegawa14}. 
The corresponding temperature-pressure $T$-$P$ phase diagram is reminiscent of those constructed for many quantum critical systems~\cite{varma99,noce00,noce02,vandermarel03,jiang09,shibauchi14,seo15} except that the magnetic transition at ambient pressure is strongly first order, although the first-order signature is significantly weakened at elevated pressures~\cite{wu14,kotegawa14,niu17}.  
At ambient pressure and low temperatures, the resistivity of CrAs follows a $T^2$ power law, supporting a Fermi-liquid behavior~\cite{wu10}. 
The Kadowaki-Woods ratio is found to be 10$^{-5} {\mu\Omega}$ cm mole$^2$ K$^2$ mJ$^{-2}$, 
which fits well to the universal value of many heavy fermion compounds~\cite{kadowaki86}. 
Nevertheless, the exponent $n$ of the low temperature behavior of resistivity, in the range between the superconducting critical temperature and T=10 K, and at the critical pressure, shows that $n\sim 1.4$~\cite{kotegawa14}, signalling a deviation from the Fermi-liquid behavior~\cite{demuer01}. 

The first order magnetic transition at the critical temperature $T_{N}$ is signalled by an abrupt changes of the lattice constants even though a lowering of the crystal structure symmetry has not been reported~\cite{shen16}. 
We point out that the interplay between structural, magnetic, and electronic properties at $T_{N}$~\cite{keller15} is rather conventional in transition metal compounds so that it is also expected to be crucial in CrAs, where the external pressure is the effective tool that settles the ground-state properties~\cite{cuoco06a,cuoco06b,cuoco10}. 

Muon spin rotation measurements performed on powder samples have revealed the existence of a region of coexistence, in the intermediate pressure region, where the superconducting and the magnetic volume fractions are spatially phase separated and compete each others~\cite{khasanov15}.
The phase separation scenario between magnetism and superconductivity together with the observation that the superfluid density  scales with the critical temperature as $\sim T_c ^{3.2}$ indicate a conventional mechanism of pairing in CrAs~\cite{khasanov15}.
This conclusion is further supported by recent transport measurements on Al-doped CrAs single crystals~\cite{park18}.  

Moreover, a nuclear quadrupole resonance under pressure~\cite{kotegawa15} shows that the internal field in the helimagnetic state decreases slowly when the pressure increases, keeping a large value close to $P_c$.
This result suggests, as previously stated, that the pressure-induced suppression of the magnetic order is of the first order. 
It has also been found that the nuclear spin-lattice relaxation rate in CrAs shows substantial magnetic fluctuations, but does not display a coherence peak in the superconducting state, supporting an unconventional pairing mechanism~\cite{kotegawa15}. 
Some direct measurements like NMR and neutron indicate that there exists strong spin fluctuations in normal state of CrAs. Indeed, there is no coherence peak at $T_c$ in NMR, and $1/T_1$ shows $T^3$ power law below T$_c$ pointing towards line-nodes in gap function. 

Very recently, it has been shown by means of neutron diffraction that coupled structural-helimagnetic order is suppressed at the pressure where the bulk superconductivity develops with the maximal transition temperature~\cite{matsuda18}.
Besides, this coupled order is also completely suppressed by phosphorus doping as in CrAs$_{1-x}$P$_x$, 
at a critical doping above which inelastic neutron scattering evidenced persistent antiferromagnetic correlations~\cite{matsuda18}. These issues provide a possible link between magnetism and superconductivity, thereby bringing out new insights towards an unconventional superconductivity similarly to that occurring in the high-T$_c$ iron pnictides.

Summarizing, at the present, no convincing picture emerges to support a conventional or an unconventional superconducting pairing in the CrAs. 

In this letter, we report detailed resistivity measurements of single crystal CrAs samples at an external pressure near to the critical one, and we investigate to what extend the normal state resistivity $\rho_{tot}(T)$ behavior is consistent with electron-phonon scattering mechanism. 
As we will show, the results give important insights on the microscopic properties of CrAs, suggesting that the CrAs cannot be considered as a phonon-mediated superconductor pointing towards other mechanisms such as, for instance, a magnetic fluctuation one as the glue driving the superconductivity.

\section{Sample preparation} The CrAs crystals were grown using Sn-flux method~\cite{wu10}. 
The starting materials were Cr (Cerac, powder, 99.9$\%$), As (Alfa Aesar, powder, 99.99$\%$), and Sn (Cerac, shot, 99.9$\%$). All of the manipulations were done in an Argon-filled glove box with moisture and oxygen levels less than 1 ppm. 
The materials with atomic ratio of Cr:As:Sn = 3:4:40 were added to an alumina crucible, which was placed in a quartz ampoule, and subsequently sealed under a reduced pressure of $10^{4}$ Torr. 
The quartz ampoule was heated up to 650 $\celsius$ for 10 h, held there for a period of 8 h, then heated up to 
 1000 $\celsius$ for 15 h, held for 6 h, and slowly cooled down to 600 ${\celsius}$ for 50 h. 
At this temperature, liquid Sn flux was filtered by centrifugation. 
The resulting products were metallic needle-shaped black crystals with dimensions up to 0.15 $\times$ 0.15 $\times$ 1 mm$^3$
The crystals were grown along the $b$-axis and they are stable in air and water. Energy-dispersive X-ray (EDX) analysis on these crystals was carried out using a Hitachi S-2700 scanning electron microscope. The results show that the chemical compositions are 51(2)$\%$ Cr, and 49(2)$\%$ As. 
No Sn atoms were detected in the crystals analyzed. The resistivity was measured between 2 and 300 K by the standard 4-probe method. The current was applied along the $b$-axis of the crystal. 

\section{ Resistivity measurements: theory and experiments} The resistivity $\rho_{tot}(T)$ of a metal can be expressed as the sum of two contributions
\begin{equation}
\label{eqn:rho}
\rho_{tot}(T)=\rho_0 + \rho(T)\, ,
\end{equation}
where $\rho_0$ is the temperature independent finite term due to the presence of defects, impurities or grain boundaries, while $\rho(T)$ accounts for the temperature-dependent electrical resistivity.
When the electrons are scattered only by phonons, without the participation of reciprocal lattice vectors, as produced by Umklapp processes, $\rho(T)$ is given by the Bloch-Gr\"{u}neisen equation~\cite{ziman60,grimvall81}. 
In the framework of this approach, the phonons that contribute to the electron-phonon interaction are the acoustic ones and the resistivity $\rho(T)$ depends only by the Debye temperature $\Theta_D$, allowing us to estimate $\Theta_D$ of the compound.  

According to the Bloch assumptions, the temperature dependent phonon contribution to the resistivity can be exactly derived. 
To this end, it is assumed that the pseudopotential form factor is constant and the lattice vibration spectra obey a Debye-like dispersion relation. Within this picture, the Umklapp processes are neglected so that the electrons couple only with longitudinal phonons. Approximating the Fermi surface and the first Brillouin zone to a sphere, one obtains the well-known Bloch-Gr\"{u}neisen result. Thus, the temperature dependent contribution to the resistivity in Eq.(\ref{eqn:rho}) may be written as
\begin{equation}
\label{eqn:rhoBG}
\rho(T)=A T^p + \rho_{BG}(T)\, .
\end{equation}

\begin{figure}[b]
	\includegraphics[width=\columnwidth]{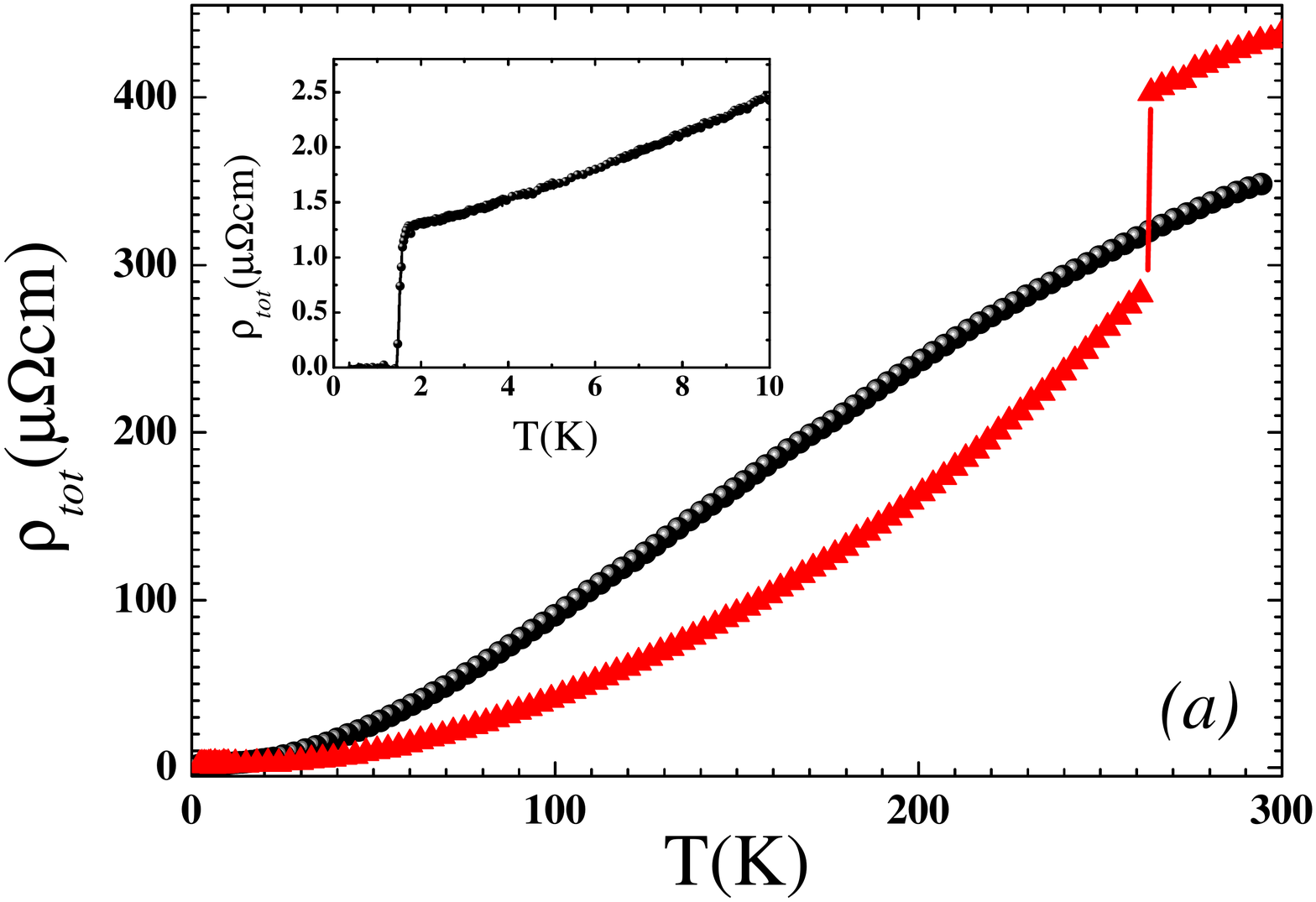}
	\includegraphics[width=\columnwidth]{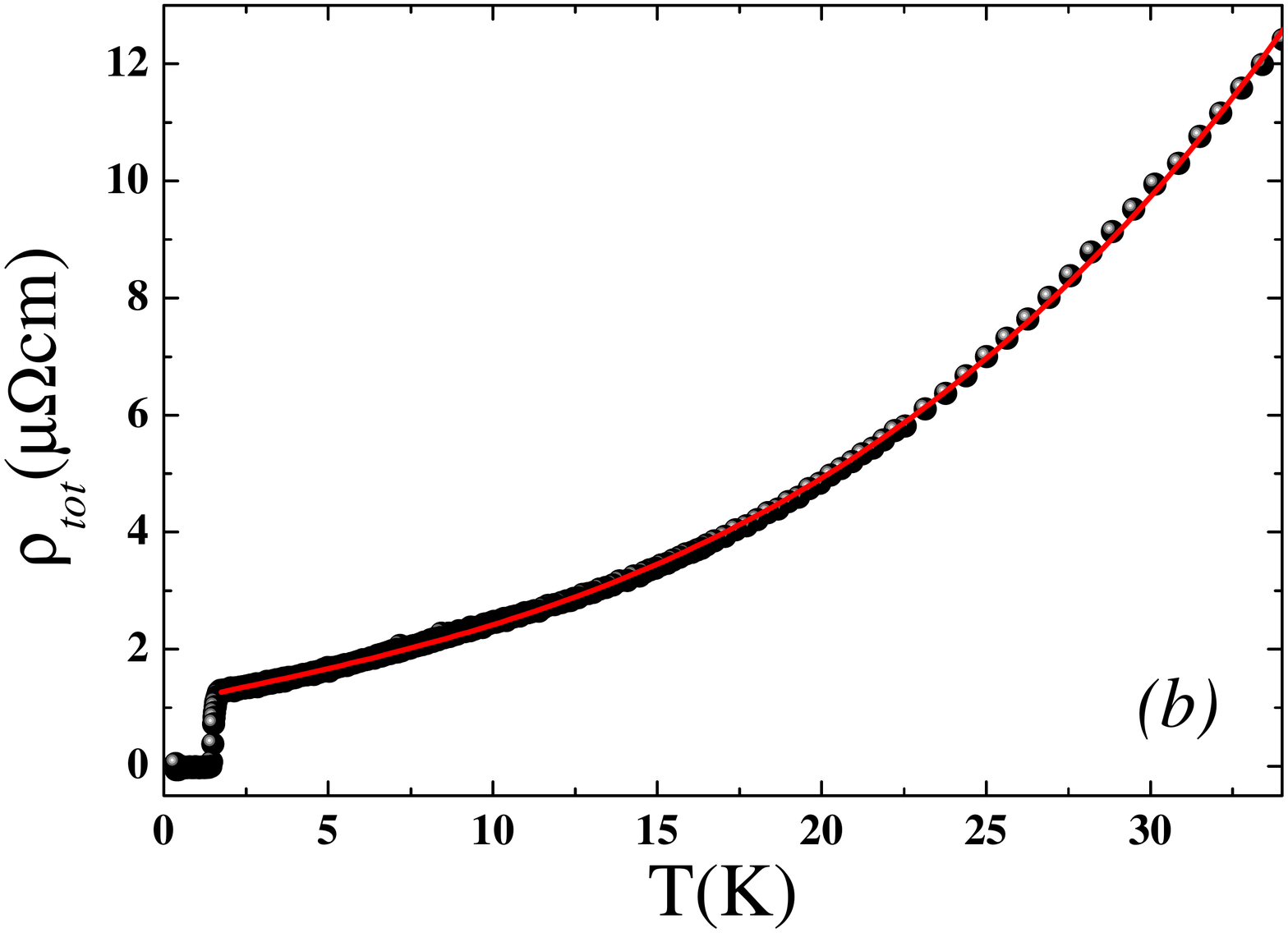}
	\caption{(Color online)
		(a) Temperature dependence of the in plane electrical resistivity $\rho_{tot}$ for a CrAs single crystal in zero magnetic field and at ambient pressure (red spheres) and at applied pressure $P=9.5~kbar$ (black squares). Inset reports the resistivity data at $P=9.5~kbar$ in the low temperature region showing the superconducting transition at $T_c$= 1.51 K.
		(b)The low temperatures resistivity data below 35 K at $P=9.5~kbar$. The red solid line is the best fit of Eq.(\ref{eqn:rhoT}).
	}
	\label{fig:1}
\end{figure}

\noindent Here, the first term accounts for the low temperature behavior of the resistivity, while the second one is the Bloch-Gr\"{u}neisen contribution. The explicit general form of $\rho_{BG}(T)$ is obtained assuming that the so-called phonon transport coupling function $\alpha_{tr}^2 F(\omega)$ may be approximated by $\sim \omega^{n-1}$\, ~\cite{grimvall81}.
This transport coupling function produces a generalized Bloch-Gr\"{u}neisen resistivity
\begin{equation}
\label{eqn:RBG}
\rho_{BG}(T)=B I^{(n)}_{BG}(T)\, ,
\end{equation}
with
\begin{equation}
\label{eqn:BG}
I_{BG}(T)= \frac{T^n}{C_n}\int_0^{{\Theta_D}/{T}}  dx
\frac{x^n}{(e^x-1)(1-e^{-x})}
\, .
\end{equation}

\noindent In Eqs.~(3)-(4), $B$ and $C_n$ are two constants, while the only relevant parameter is the Debye temperature $\Theta_D$, allowing us to estimate $\Theta_D$ of the compound.  
For completeness, we notice that if  $\alpha_{tr}^2 F(\omega)\propto\omega^4$, one
gets the well-known $T^5$ power law~\cite{ziman60}.

At low temperatures, that is for $T\ll\Theta_D$, the resistivity as defined in Eqs.(\ref{eqn:rho})-(\ref{eqn:rhoBG}) can be modelled through a temperature power law which reads as
\begin{equation}
\label{eqn:rhoT}
\rho_{tot}(T)=\rho_0 + A T^p + B T^n\, ,
\end{equation}
where the Bloch-Gr\"{u}neisen term $I^{(n)}_{BG}(T)$ approximates to $T^n$, as derived from Eq.(\ref{eqn:BG}).

\begin{figure}
	\includegraphics[width=\columnwidth]{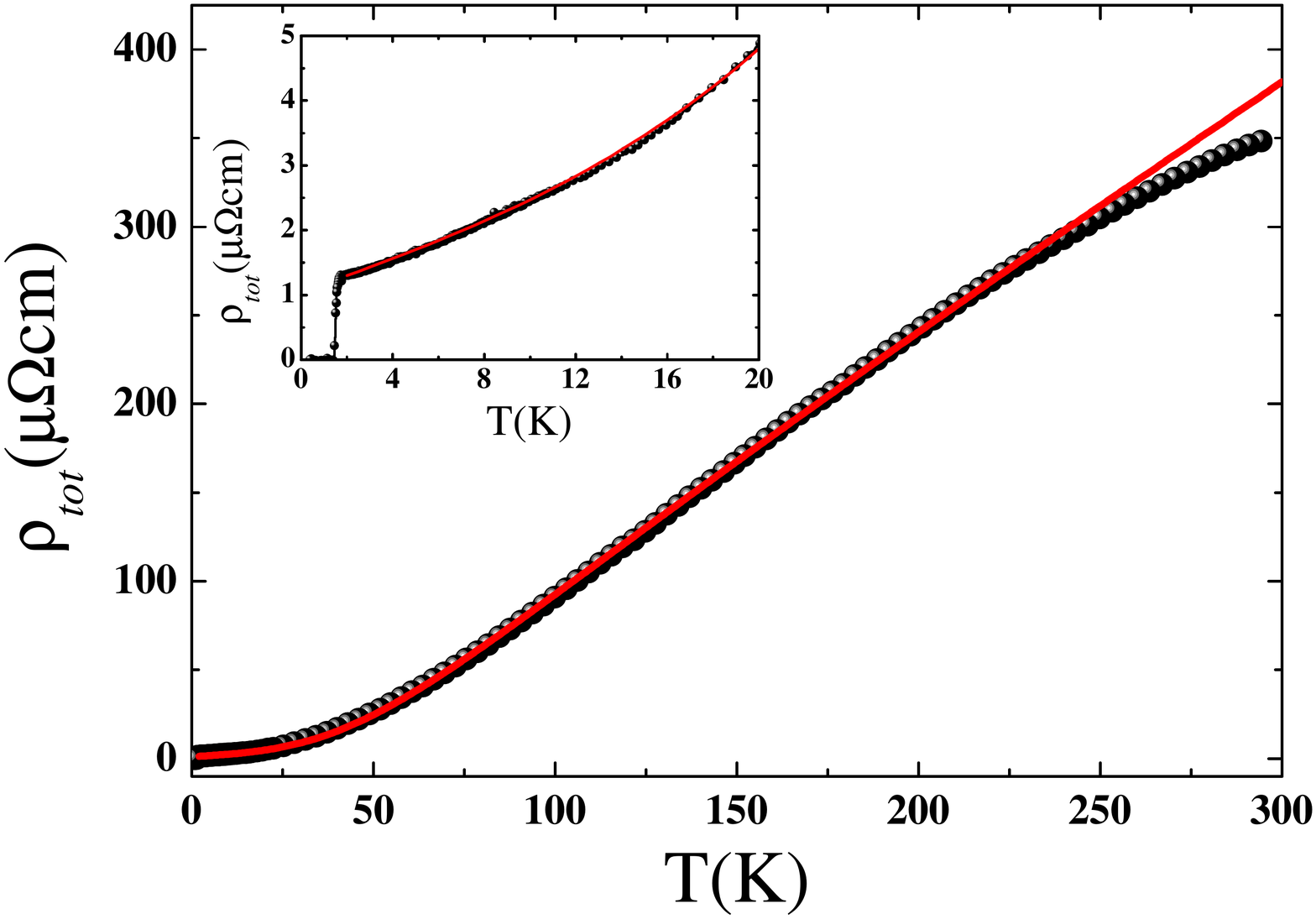}
	\caption{(Color online)
		The best fit (red solid line) of $\rho_{tot}(T)$ as given by equations (1) and (2) to the experimental data (black spheres), in the temperature range 2--250~K. The inset shows an enlarged view below 20 K.}
	\label{fig:2}
\end{figure}

Fig.(\ref{fig:1}a) shows the temperature dependence of the in plane electrical resistivity $\rho_{tot}(T)$ for a selected CrAs single crystal sample at ambient pressure and in zero magnetic field. We observe that $\rho_{tot}(T)$ (red triangles) exhibits a typical metallic behavior over the reported temperature range.  
A drop in the electrical resistivity is observed at $T_N$=270 K indicating the presence of a magnetic transition. This magnetic transition occurs with a simultaneous structural distortion with the lattice parameter $b$ increasing abruptly by $4\%$, while $a$ and $c$ decrease by less than $1\%$.

In the same Fig.(\ref{fig:1}a) it is also reported $\rho_{tot}(T)$ (black circles) at the pressure $P=$9.5~kbar.
First of all, we notice that a residual resistivity $\rho_0$ equal to ${1.2}{\mu \Omega cm}$ is measured, and the large residual-resistivity-ratio, $RRR=\rho(300{K})/\rho_0\,\simeq300$, confirms the high quality of the used CrAs single crystal. At this pressure, we observe no anomaly related to the magnetic transition, whereas a single step superconducting transition is visible at $T_c$=1.51 K with a transition width $\Delta T_c$ =0.12 K, as shown in the inset of Fig.(\ref{fig:1} a).
We point out that $T_c$ has been evaluated as the temperature corresponding to 50$\%$ of the normal state resistivity at the onset of the superconducting transition $\rho_n$, while $\Delta T_c$ is defined as the difference between the temperature corresponding to 90$\%$ and 10$\%$ of resistivity at the superconducting transition temperature drop.

Fig.(\ref{fig:1}b) shows the low temperatures resistivity data below 35 K.
The red solid line is the best fit of Eq.(\ref{eqn:rhoT}).
The data fitting yields $p=1$, $n=3$ and a residual resistivity $\rho_0={1.1}{\mu\Omega cm}$.
Previous studies showed the low temperature resistivity of many transition metal alloys~\cite{jiang15}
is well described by an exponent $n=3$ rather than $n=5$.
In this case the electron scattering from $s$ to $d$ bands and/or electron-magnon scattering occurs giving rise to this temperature power law resistivity behaviour~\cite{white59,mott64}. This result suggests that the phonon-mediated contribution is not able to reproduce the temperature trend of the resistivity. 

In the temperature range [2;250]~K, a best fit of $\rho_{tot}(T)$ as given by equations (1) and (2), with $n = 3$, with $C_3 = 7.212$, and $p=1$, to the experimental data has been carried out and shown in Fig.(\ref{fig:2}) (red solid line). 
The Debye temperature $\Theta_D$, the coefficients $A$ and $B$ are considered as free fitting parameters:
The best fit values we get are $\Theta_D = (340\pm 5) \, K $, $B = (14.7 \pm 0.2) 10^{-5} \,{\mu \, \Omega\,  cm\,  K^{-3}}$ and $A = (0.13\pm 0.01)\, {\mu\,  \Omega\,  cm\,  K^{-1}}$. 

We point out that a deviation from the fitting model of the temperature behavior of resistivity  is  observed at high temperatures where the resistivity seems to saturate. In particular, a decrease in the slope $d\rho/dT$ with increasing temperature is clearly recognized above $\simeq$ 240 K in Fig.(\ref{fig:2}).

\section{ Final remarks}  We have shown that the temperature trend of $\rho_{tot}(T)$ cannot be explained by the Bloch-Gr\"{u}neisen contribution only since it is not enough to reproduce the experimental data, but we need an extra term accounting for non-phononic contributions. We point out that a similar temperature behaviour has been traced for BaVSe$_3$ and ascribed to electron-electron scattering. This mechanism dominates other scattering mechanisms, such as for instance the electron-phonon interaction, showing up a deviation from the $n$=2 power law exponent of the temperature dependent resistivity~\cite{akrap08}.

Moreover, it is well-known that the resistivity measurements may gives an estimation of the electron-phonon coupling strength relevant for superconductivity~\cite{allen87,allen99}. Thus, considering the fitting of the temperature behaviour of the resistivity, we may derive a rough estimation of the average electron phonon-coupling constant $\lambda_{e-ph}$. 
Picking up the measured value of the critical temperature $T_c$=1.5 K and the estimated Debye temperature $\Theta_D$= 340 K, through the BCS equation
\begin{equation}
T_c=1.14 \Theta_D \exp(-1/\lambda_{e-ph})\, ,
\end{equation}
we get $\lambda_{e-ph}\sim 0.18$.
This value of $\lambda_{e-ph}$ is very low so that it suggests a very weak coupling regime, hence supporting the hypothesis that the electron-phonon coupling plays a minor role in the formation of Cooper pairs~\cite{allen87,allen99}. 

Furthermore, in the framework of Eliashberg theory~\cite{eliashberg60}, the Allen-Dynes formula~\cite{allen75} yields an estimate of the critical temperature, which accounts for the competition of the phonon driven attractive interaction with the repulsive Coulomb interaction expressed by the parameter $\mu^*$. This parameter describes the Coulomb repulsion reduced by retardation and screening effects and is obtained using the formula given by Morel and Anderson for the retarded Coulomb potential~\cite{morel62}. For conventional superconductors retardation effects are believed to be very important in view of the very different energy scales for the electrons and phonons. From the available data on the density of states of CrAs~\cite{autieri17a,autieri17b,autieri18,cuono18}, we are able to estimate $\mu^*$~\cite{nixon07}, obtaining $\mu^*$= 0.12. Since this value is not very much reduced by retardation effects, and considering that an efficient screening is important in reducing this parameter sufficiently to allow for an electron-phonon driven superconductivity~\cite{koch99,bauer13}, we may conclude that CrAs differ from the conventional phonon picture, where retardation effects play a major role in reducing the electron-electron repulsion.

Therefore, even though no clear indication on the nature of the superconducting states is currently available, from our results we infer that the extremely low electron-phonon coupling constant and the not strongly renormalized Coulomb screening parameter indicate that a non-phonon-mediated superconducting coupling is likely, hence supporting an electronic driven mechanism such as, for instance a magnetic-mediating coupling.

\acknowledgments
The work is supported by the Foundation for Polish Science through the IRA Programme co-financed by EU within SG OP.

\end{document}